\documentclass[runningheads]{llncs}

\usepackage{graphicx}
\usepackage{subcaption}
\usepackage{amsfonts} 
\usepackage{amsmath}
\usepackage{amsfonts}
\usepackage{bbold}
\usepackage{booktabs}
\usepackage{hyperref}
\usepackage{fontawesome}

\usepackage{listings}
\begin{document}
\title{TMSS: An End-to-End Transformer-based Multimodal Network for Segmentation and Survival Prediction \thanks{This work has a filed patent in the US (USPTO: 17849943)}}
\titlerunning{TMSS}
\author{Numan Saeed$^{0000-0002-6326-6434}$ \textsuperscript{\faEnvelopeO} \and
Ikboljon Sobirov$^{0000-0002-0476-6359}$\and
Roba Al Majzoub$^{0000-0002-8349-8953}$\and 
Mohammad Yaqub$^{0000-0001-6896-1105}$ \textsuperscript{\faEnvelopeO}
}

%index {Saeed, Numan} {Sobirov, Ikboljon} {Al Majzoub, Roba} {Yaqub, Mohammad}

\authorrunning{Numan Saeed et al.}

\institute{
Mohamed bin Zayed University of Artificial Intelligence, Abu Dhabi, UAE 
\email{\{numan.saeed, ikboljon.sobirov, roba.majzoub, mohammad.yaqub\}@mbzuai.ac.ae}
}
\maketitle              % typeset the header of the contribution
\begin{abstract}
When oncologists estimate cancer patient survival, they rely on multimodal data. Even though some multimodal deep learning methods have been proposed in the literature, the majority rely on having two or more independent networks that share knowledge at a later stage in the overall model. On the other hand, oncologists do not do this in their analysis but rather fuse the information in their brain from multiple sources such as medical images and patient history. This work proposes a deep learning method that mimics oncologists' analytical behavior when quantifying cancer and estimating patient survival. We propose TMSS, an end-to-end \textbf{T}ransformer based \textbf{M}ultimodal network for \textbf{S}egmentation and \textbf{S}urvival predication that leverages the superiority of transformers that lies in their abilities to handle different modalities. The model was trained and validated for segmentation and prognosis tasks on the training dataset from the HEad \& NeCK TumOR segmentation and the outcome prediction in PET/CT images challenge (HECKTOR). We show that the proposed prognostic model significantly outperforms state-of-the-art methods with a concordance index of $\textbf{0.763} \pm{\textbf{0.14}}$ while achieving a comparable dice score of $\textbf{0.772} \pm{\textbf{0.030}}$ to a standalone segmentation model. The code is publicly available at \url{https://t.ly/V-_W}.

\end{abstract}
\section{Introduction}

Cancer is one of the most lethal diseases and is among the top leading causes of death as reported by the World Health Organization~\cite{WHOCancer} with almost 10 million deaths in 2020. This fact encourages doctors and medical researchers to strive for more efficient treatments and better care for cancer patients. Head and Neck (H\&N) cancer is a collective term used to describe malignant tumors that develop in the mouth, nose, throat, or other head and neck areas. Early prediction and accurate diagnosis of the patient's survival risk (prognosis) can lower the mortality rate to 70\%~\cite{wang} of the H\&N cancer patient. Furthermore, an accurate prognosis helps doctors better plan the treatments of patients~\cite{impotanceProg}. Doctors routinely conduct different types of scans like computed tomography (CT) and positron emission tomography (PET) in clinics and utilize them to extract biomarkers of the tumor area that is used with other information like patients electronic health records (EHR)for treatment plans. However, the manual delineation of the scans is time-consuming and tedious. Automatic prognosis and segmentation can significantly influence the treatment plan by speeding up the process and achieving robust outcomes. 

% Many attempts to automate prognosis and enhance its results through different Machine Learning (ML) methods, specifically Deep Learning (DL), have created customized treatment plans for each patient. However, none, to the best of our knowledge, of the current systems utilize multimodal data of patients in such an end-to-end manner to segment and prognose tumors, as oncologists do. Instead, the current state-of-the-art methods use deep neural networks to extract features from image scans, concatenate them with the EHR (tabular) data and feed the concatenated data to survival models for risk score predictions~\cite{saeed2022ensemble}.

With the advent of the deep learning (DL) field, considerable effort has been put into the automatic prognostic analysis for cancer patients. Brain~\cite{zhou,brain}, breast~\cite{breast,breast2}, liver~\cite{liver,liver2,liver3}, lung~\cite{lung,lung2}, rectal~\cite{rectal} and many other cancer types have been previously studied extensively. Long-term survival prediction using 33 different types of cancer was examined in-depth in~\cite{vale}. Their MultiSurv multimodal network is compromised of several submodules responsible for feature extraction, representation fusion, and prediction. A multimodal deep neural network using gene expression profile, copy-number alteration profile, and clinical data was proposed in~\cite{breast} for breast cancer prognosis. In~\cite{liver2}, an improvement on the prognosis of patients with colorectal cancer liver metastases was studied. The authors proposed an end-to-end autoencoder neural network for this task utilizing radiomics features taken from MRI images. For overall survival prediction of patients with brain cancer, authors in ~\cite{zhou} proposed an end-to-end model that extracts features from MRI images, fuses them, and combines outputs of modality-specific submodels to produce the survival prediction.

Using DL to predict a patient's outcome with H\&N cancers is understudied. Clinically, H\&N squamous cell carcinoma refers to different types of H\&N cancers~\cite{johnson}, including our topic of interest - oropharynx cancer. Authors in~\cite{diamant} studied H\&N squamous cell carcinoma, creating an end-to-end network and arguing that a basic CNN-based model can extract more informative radiomics features from CT scans to predict H\&N cancer treatment outcomes. H\&N squamous cell carcinoma prognosis and its recurrence using DL were examined in~\cite{fh}. The authors used CT scans of patients diagnosed with this type of cancer and extracted radiomics features manually using gross tumor volume and planning target volume. They predicted H\&N cancer-related death and recurrence of cancer using a DL-driven model. Oropharyngeal squamous cell carcinoma, in particular, was a topic of interest in~\cite{fujima}. PET scans were used to train different popular CNN architectures, such as AlexNet~\cite{alexnet}, GoogleLeNet~\cite{goglenet}, and ResNet~\cite{resnet}, all of which were pretrained on ImageNet~\cite{imagenet}, to compare it with the traditional methods that are trained on clinical records. By comparing all four different approaches, they concluded that using PET scans for a diagnostic DL model can predict progression-free survival and treatment outcome.

In~\cite{saeed2022ensemble}, authors tackled the prognosis task for oropharyngeal squamous cell carcinoma patients using CT and PET images, along with clinical data. Their proposed solution was ranked the first in the progression-free survival prediction task of MICCAI 2021 HEad and neCK TumOR (HECKTOR) segmentation and outcome prediction challenge~\cite{hecktor}. Features from medical images were first extracted using a CNN-based module and concatenated with electronic health records. The outputs were passed through fully connected layers and then to a multi-task logistic regression (MTLR) model~\cite{mtlr}. Parallelly, electronic health records were fed to a Cox proportional hazard (CoxPH) model~\cite{cox} to predict the patients' risk score. Finally, the risk predictions were calculated by taking the average of the outputs from MTLR and CoxPH models. This ensemble model achieved a concordance index (C-index; a common metric to evaluate prognosis accuracy) of 0.72 on the HECKTOR testing set, outperforming other proposed solutions. Although this work utilizes multimodal data, the learning from the medical images and EHR were disjoint, which may lead to less discriminative features learned by the deep learning model and consequently affect the final outcome.

In this paper, we propose a novel end-to-end architecture, TMSS, to predict the segmentation mask of the tumor and the patient's survival risk score by combining CT, PET scans and the EHR of the patient. Standard convolutional neural networks mainly focus on the imaging modality and cannot use other input features. To generally address this concern, we propose a transformer-based encoder that is capable of attending to the available multimodal input data and the interaction between them.  

In the current work, our contributions are as follows: 

\begin{itemize}
    \item We propose \textit{TMSS}, a novel end-to-end solution for H\&N cancer segmentation and risk prediction. \textit{TMSS} outperforms the SOTA models trained on the same dataset.
    \item We show that a vision transformer encoder could attend to multimodal data to predict segmentation and disease outcome, where the multimodal data is projected to the same embedding space.
    \item We propose a combined loss function for segmentation mask and risk score predictions.
\end{itemize}

\section{Proposed Method}
% In the following section, the build up of the main architecture is described. The architecture at a glance is composed of several components, including embedding and positional encoding for images and EHR, fusion of the multimodal data, encoder, decoder for segmentation end, and DeepMTLR for prognostic end.  
% where we start with EHR data and shift afterwards to combine the two modalities and therefore the architecture will evolve to accommodate both modalities. Note that all models underwent KFold cross validation with K=5.

\begin{figure}[htbp]\centering
{\includegraphics[width=\linewidth]{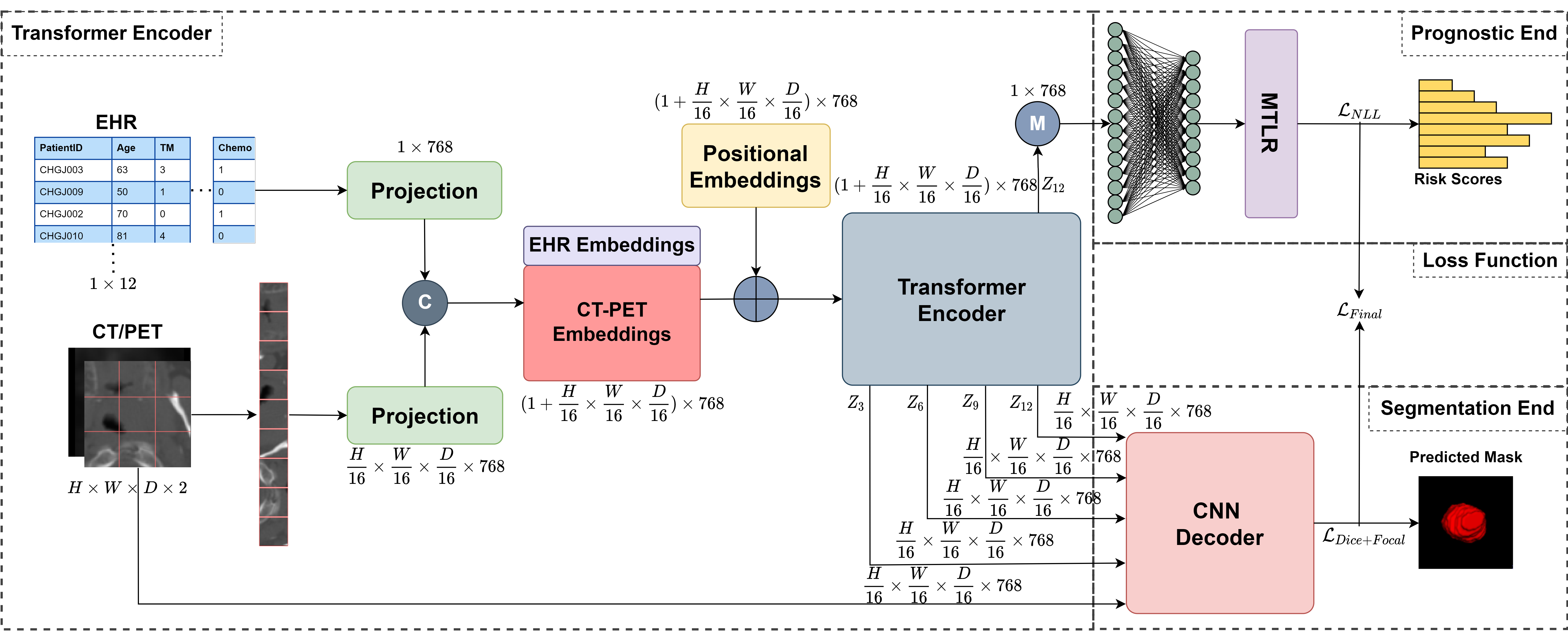}}
\caption{An illustration of the proposed TMSS architecture and the multimodal training strategy. TMSS linearly projects EHR and multimodal images into a feature vector and feeds it into a Transformer encoder. The CNN decoder is fed with the input images, skip connection outputs at different layers, and the final layer output to perform the segmentation, whereas the prognostic end utilizes the output of the last layer of the encoder to predict the risk score.}
\label{architecture}
\end{figure}

In the following section, we describe the build up of the main architecture, depicted in Figure~\ref{architecture}. As can be seen, it comprises four major components, and each is further explained below. 

\subsubsection{Transformer Encoder.}
The main advantage of this network is that the encoder itself embeds both the CT/PET and EHR data and encodes positions for them accordingly while extracting dependencies (i.e. \textbf{attention}) between the different modalities. The 3D image with dimensions $x\in \mathbb{R}^{H\times W\times D\times C}$ is reshaped into a sequence of flattened 2D patches $x_p\in \mathbb{R}^{n\times (P^3C)}$, where $H$, $W$, and $D$ are the height, width, and depth of the 3D image respectively, $C$ denotes the number of channels, $P\times P\times P$ represents each patch's dimensions, and $n=HWD/P^3$ is the number of patches extracted. These patches are then projected to the embedding dimension $h$, forming a matrix $I\in \mathbb{R}^{n\times h}$. Simultaneously, EHR data is also projected to a dimension $E\in \mathbb{R}^{1\times h}$. Both projections of images and  EHR are concatenated, forming a matrix $X\in \mathbb{R}^{(n+1)\times h}$. Positional encodings with the same dimension are added to each of the patches and the EHR projection as learnable parameters. The class token is dropped from the ViT \cite{vit} as our solution does not address a classification task. The resulting embeddings are fed to a transformer encoder consisting of 12 layers, following the same pipeline as the original ViT, with normalization, multi-head attention, and multi-layer perceptron. The purpose of using self-attention is to learn relations between $n+1$ number of embeddings, including images and EHR. The self-attention inside the multi-head attention can be written as \cite{vit}:

\begin{equation}
\label{attention}
    Z = softmax\left(\dfrac{QK^T}{\sqrt{D_q}}\right)V
\end{equation}

% The patch embedding for the imaging data is similar to that of original ViT model, whereas EHR is projected to the hidden size. 

% $B\times n\times h$, where $B$ is the batch size, $n$ is the number of patches, and $h$ is the hidden size, are the embedding dimensions for the imaging data. The EHR data is with the size of $B\times 1\times h$ after its embedding, and is concatenated with embeddings of images, creating the size $B\times (n+1)\times h$. Positional encoding of the same size are added 

\subsubsection{Segmentation End.}
The segmentation end is a CNN-based decoder, similar to the decoder in~\cite{UNETR}. The original images are fed to the decoder along with skip connections passed from ViT layers $Z_3$, $Z_6$, $Z_9$, and $Z_{12}$ (last layer). Only the image latent representations are passed through these skip connections  $Z_{l}\in \mathbb{R}^{(n)\times h}$ and fed to the CNN decoder, where $l \in \{ 3, 6, 9, 12 \}$. Convolution, deconvolution, batch normalization, and Rectified Linear Unit (ReLU) activation are used in the upsampling stage. Please, refer to~\cite{sobirov2022automatic} for more details. 

\subsubsection{Prognostic End.}
The prognostic path receives the output of the encoder with dimensions $Z_{12}\in \mathbb{R}^{(n+1)\times h}$, and its mean value is computed, reducing the dimensions down to $Z_{mean}\in \mathbb{R}^{1\times h}$. This latent vector is then forwarded to two fully connected layers, reducing the dimensions from $h$ to $512$ and $128$ respectively. The resulting feature map is then fed to an MTLR model for final risk prediction. The MTLR module divides the future horizon into different time bins, set as a hyperparameter, and for each time bin a logistic regression model is used to predict if an event occurs or not.  

\subsubsection{Loss Function.}
Since the network performs two tasks concurrently, a combination of three losses is formulated as the final objective function. The segmentation end is supported by the sum of a dice loss (Equation~\ref{dice_loss}) and a focal loss (Equation~\ref{focal_loss}), where $N$ is the sample size, $\hat{p}$ is the model prediction, $y$ is the ground truth, $\alpha$ is the weightage for the trade-off between precision and recall in the focal loss (set to 1), and $\gamma$ is focusing parameter (empirically set to 2).

\begin{equation}
   \label{dice_loss}
   \mathcal{L}_{Dice} = \frac{2\sum_{i}^{N} \hat{p_i} y_i}{\sum_{i}^{N} \hat{p_i}^2 + \sum_{i}^{N} y_i^2},
\end{equation}

\begin{equation}
   \label{focal_loss}
   \mathcal{L}_{Focal} = -\sum_{i}^{N}\alpha y_i (1 - \hat{p_i})^{\gamma}log(\hat{p_i}) - (1 - y_i)\hat{p_i}^{\gamma}log(1-\hat{p_i}),
\end{equation}

The prognostic end has a negative-log likelihood loss (NLL) as given in Equation~\ref{nnl_loss}. Here, the first line in the NLL loss corresponds to uncensored data, the second line corresponds to censored data and the third line is the normalizing constant, as described in~\cite{kazmierskithesis}. The product $w_k^Tx^{(n)}$ is the model prediction, $b_k$ is the bias term, and $y_k$ is the ground truth.   

\begin{equation}
   \label{nnl_loss}
  \begin{split}
        &\mathcal{L}_{NLL}(\theta, D) = \sum_{n: \delta^{(n)}=1} \sum_{k=1}^{K-1}(w_k^Tx^{(n)} + b_k)y_k^{(n)} \\
        &+ \sum_{n: \delta^{(n)}=0}  log\left(\sum_{i=1}^{K-1}\mathbb{1}\{t_i \geq T^{(n)}\}exp\left(\sum_{k=1}^{K-1}\left((w_k^Tx^{(n)}+b_k)y_k^{(n)} \right) \right) \right) \\
       &- \sum_{n=1}^{N}log\left(\sum_{i=1}^{K}exp\left(\sum_{k=1}^{K-1}w_k^Tx^{(n)} + b_k \right) \right) ,
    \end{split}
\end{equation}

The final loss used for our network training is provided in Equation~\ref{final_loss} as a combination of the three losses. The hyperparameter $\beta$, provides weightage to either side of the model paths, and is empirically set to 0.3.

\begin{equation}
   \label{final_loss}
   \mathcal{L}_{Final} = \beta * (\mathcal{L}_{Dice} + \mathcal{L}_{Focal}) + (1 - \beta) * \mathcal{L}_{NLL}
\end{equation}

\section{Experimental Setup}

\begin{figure}[htbp]
\captionsetup[subfigure]{justification=centering}
\centering
\begin{minipage}{0.24\textwidth}
\begin{subfigure}{\textwidth}
    \includegraphics[width=\textwidth]{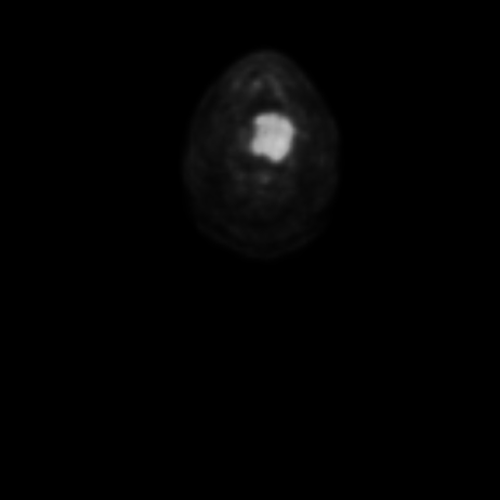}
    \subcaption{\textbf{Original PET}}
\end{subfigure}
\end{minipage}
\begin{minipage}{0.24\textwidth}
\begin{subfigure}{\textwidth}
    \includegraphics[width=\textwidth]{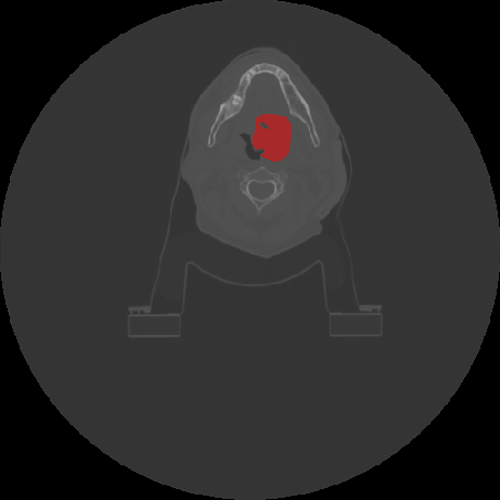}
    \subcaption{\textbf{Original CT}}
\end{subfigure}
\end{minipage}
\begin{minipage}{0.24\textwidth}
\begin{subfigure}{\textwidth}
    \includegraphics[width=\textwidth]{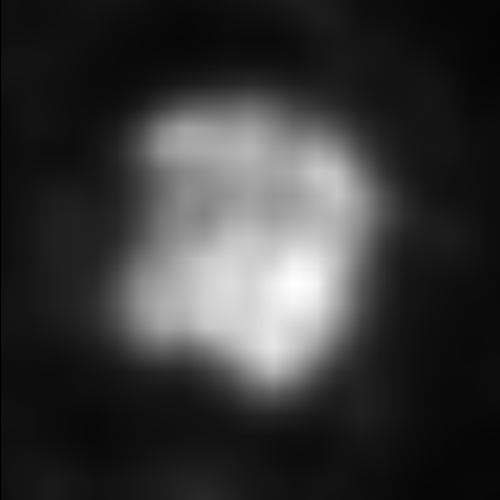}
    \subcaption{\textbf{Cropped PET}}
\end{subfigure}
\end{minipage}
\begin{minipage}{0.24\textwidth}
\begin{subfigure}{\textwidth}
    \includegraphics[width=\textwidth]{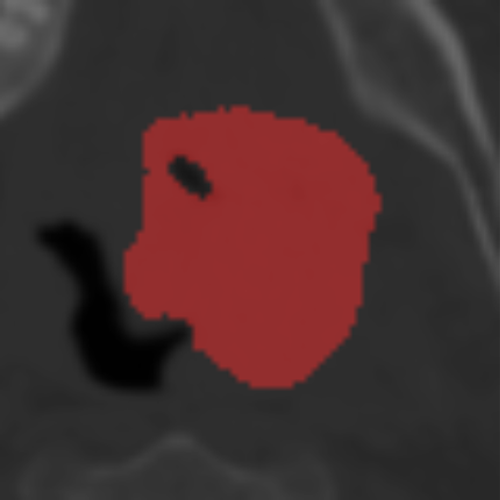}
    \subcaption{\textbf{Cropped CT}}
\end{subfigure}
\end{minipage}
\caption{A sample from the imaging dataset. (a) depicts the original PET scan (b) depicts the original CT scan and the imposed ground truth mask (c) shows the $80\times80\times48$ cropped PET and (d) shows the $80\times80\times48$ cropped CT with ground truth mask.}
\label{data_sample}
\end{figure}
\subsection{Dataset Description}
A multicentric dataset of PET and CT images, their segmentation masks, and electronic health records are available on the HECKTOR challenge platform\footnote[1]{aicrowd.com/challenges/miccai-2021-hecktor}. The data comes from six different clinical centers; 224 and 101 patient records for training and testing respectively. The testing set ground truths, both for segmentation and prognosis tasks are hidden for competition purposes, thus are not used to validate our method. Therefore, \emph{k}-fold (where \emph{k}=5) cross validation was performed on the training set. EHR is comprised of data pertinent to gender, weight, age, tumor stage (N-, M- and T-stage), tobacco and alcohol consumption, chemotherapy experience, human papillomavirus (HPV), TNM edition, and TNM group. Imaging data contains CT, PET, and segmentation masks for tumor; sample slices are illustrated in Figure~\ref{data_sample} respectively. 

% \subsection{Tumor Localization}
% In \cite{sobirov2022automatic}, we addressed the first task of segmentation of H\&N tumor of the HECKTOR challenge. We used a transformer-based model for segmentation. For the CT and PET images, the initial preprocessing was to use the predicted segmentation masks from \cite{sobirov2022automatic}. Specifically, the tumor center was calculated, and the original scans were cropped down to the size of 80$\times$80$\times$48 around the tumor center, as depicted in Figure~\ref{data_sample}. This significantly reduces the image sizes and accentuates the tumor region, assisting our model to learn more easily and making it computationally superior. 

\subsection{Data Preprocessing}
Both the CT and PET images are resampled to an isotropic voxel spacing of 1.0\emph{$mm^3$}. Their intensity values are then normalized before being fed to the network. The window of HU values of CT images was empirically clipped to (-1024, 1024), after which the images were normalized between (-1, 1). On the other hand, Z-score normalization was used for PET images. Furthermore, the images are cropped down to $80\times80\times48mm^3$ as in \cite{saeed2022ensemble} for two main purposes; the first is to fairly compare our results to the state-of-the-art in \cite{saeed2022ensemble}, which also used images with these dimensions. The second is that this reduction of image dimensions, in turn, speeds up training and inference processes and allows to run multiple experiments.  

EHR, being multicentric, is missing some data for tobacco, alcohol consumption, performance status, HPV status, and estimated weight for SUV from most of the centers; therefore, they were dropped. 75\% of the total data is censored, assumed to have stopped the follow-up to the hospitals. 

\subsection{Implementation Details}
For our experiments, we used a single NVIDIA RTX A6000 (48GB). We used PyTorch to implement the network and trained the model for 50 epochs. The batch size was set to 16, the learning rate to 4e-3, and the weight decay to 1e-5. The step decay learning rate strategy was used to reduce the learning rate by a factor of 10 after the 35 epochs. 

% \begin{table}
% \centering
% \caption{Prognosis performance by different models on the HECKTOR dataset. The reported are the mean and standard deviation for 5-fold cross validation.}
% \label{ci_results}
% \setlength{\tabcolsep}{6.5pt}
% \begin{tabular}{cccccc}

% \hline
% & CoxPH& MTLR& Deep MTLR& Ensemble~\cite{saeed2022ensemble}& \textbf{Ours}  \\

% \hline
% C-index & 0.682\scriptsize\textpm 0.06 & 0.600\scriptsize\textpm 0.031& 0.692\scriptsize\textpm 0.06 & 0.704\scriptsize\textpm 0.07 & \textbf{0.763}\textbf{\scriptsize\textpm 0.14} \\

% \hline
% \end{tabular}
% \end{table}

\begin{table}[]
    \centering
    \caption{Prognosis performance by different models on the HECKTOR dataset. The reported are the mean and standard deviation for 5-fold cross validation.}
    \label{ci_results}
    \setlength{\tabcolsep}{6.5pt}
    \begin{tabular}{lcccccl}\toprule[1.5pt]
         & CoxPH& MTLR& Deep MTLR& Ensemble~\cite{saeed2022ensemble}& \textbf{Ours}\\\midrule
         C-index & 0.682\scriptsize\textpm 0.06 & 0.600\scriptsize\textpm 0.031& 0.692\scriptsize\textpm 0.06 & 0.704\scriptsize\textpm 0.07 & \textbf{0.763}\textbf{\scriptsize\textpm 0.14} \\\bottomrule[1.5pt]
    \end{tabular}

\end{table}

The scans are patched into the size of $16\times16\times16$, and projected to the embedding dimension of 768. The total number of layers used in the encoder was 12, each having 12 attention heads.

% CT and PET scans were passed to the transformer encoder as two channels. The scans are patched into the size of $16\times16\times16$, and projected to the embedding dimension of 768. The total number of layers used in the encoder was 12, each having 12 of attention heads. 

% The prognostic end has two fully connected (FC) layers and the MTLR module. The dropout was set to 0.2 in the FC layers, each layer decreasing the transformer output dimension to 512 and 128 respectively. MTLR divides the future horizon into 16 time bins. 

The $\beta$ in the loss function was set to 0.3. All the hyperparameters were chosen empirically, using the framework \texttt{OPTUNA}~\cite{optuna}. The evaluation metrics for the prognosis risk was concordance index (C-index), and for the segmentation was dice similarity coefficient (DSC).

\section{Experimental Results}
We use the HECKTOR dataset as described above for the diagnosis and prognosis of patients with head and neck cancer. Several experiments were conducted in house, all in $5$-fold cross validation using the training dataset from the challenge. All the experiments were trained and cross validated using the same settings.

Table~\ref{ci_results} shows the results of all conducted experiments. We started with the commonly used algorithms for survival analysis. CoxPH, MTLR and Deep MTLR were applied as baselines. As is vivid, CoxPH, achieving C-index of 0.68, outperforms the MTLR model by a huge degree of 0.08, yet introducing neural nets to MTLR (i.e. Deep MTLR) boosted the score to 0.692. All three calculate the risk using only the EHR data on account of their architectural nature. An ensemble of CNNs for the images with MTLR, and CoxPH for EHR achieved the highest C-index on the testing set \cite{saeed2022ensemble} which was also implemented to train and validate using the same fashion as the original work. The ensemble was able to reach C-index of 0.704. Finally, our model, embedding EHR information in the input and using transformers unlike the ensemble, outperforms all the other models, achieving a mean C-index of \textbf{0.763}. 

We optimized the hyperparameters using only one of the folds and then performed k-fold cross-validation using the entire dataset. However, there is a chance of leakage due to that specific fold; therefore, we redid the testing using a hold-out test set. To rule out the statistical dependence of the model training, we split the dataset randomly into two subsets, train and test, with an 80\% and 20\% ratio, respectively. The model hyperparameters were optimized using a small subset of the training set and tested using the hold-out set. We got a C-index score of \textbf{0.74} on the testing set. The prognosis task score on the hold-out set is slightly lower than the k-fold cross-validation score of 0.76, but it is more reliable and greater than the previous best scores of 0.70. 

For segmentation comparison purposes we implement UNETR, a segmentation standalone network using the same settings as \cite{sobirov2022automatic}. Our model achieved DSC of $0.772 \pm{0.03}$, which was only 0.002 lower than that of UNETR network optimized for segmentation which achieved DSC of 0.774\textpm0.01.

% \begin{table}
% \centering
% \caption{Segmentation performance by the proposed model on the HECKTOR dataset.}
% \label{dice_results}
% \setlength{\tabcolsep}{5pt}
% \begin{tabular}{|c|c|c|}
% \hline
% & UNETR Standalone & Ours  \\

% \hline
% DSC & 0.774\scriptsize\textpm 0.01 & 0.772\scriptsize\textpm 0.03\\

% \hline
% \end{tabular}
% \end{table}

\section{Discussion}
The traditional approach to automate  diagnosis and prognosis of cancerous patients is generally performed in two stages; either a standalone network that extracts tumor radiomics such as tumor volume \cite{kazmierskithesis} and feeds it to a prognostic model, or as in SOTA~\cite{saeed2022ensemble} that uses an ensemble of CNNs to extract scans features and concatenate with the EHR, then feeds them to another network for the risk prediction.

However, our approach tackles both problems at once, in an end-to-end network, making it simpler and easier to train. We show how our method outperforms other models by a good margin using vision transformers. Encoding EHR data into the network was newly introduced to mimic the way doctors review patient data. This has effectively boosted the accuracy of prognosis as shown in Table \ref{ci_results}. The aforementioned results show the superiority of transformers in handling multimodal data. We hypothesize that the attention embedded in the transformer blocks, along with their ability to accommodate multimodal data, allows them to find relations across the modalities and within their intermediate representations. That can help them better address the tasks at hand. The use of multiple losses boosts the ability of the model to better interpolate within the given data, and hopefully become more robust when subjected to unseen ones. The introduction of the weighting variable $\beta$ with a value of 0.3 penalizes the model more for prognosis errors coercing it to learn the features better and adjust its weights accordingly for an accurate prognosis. 

Although the main goal of our model is prognosis, and not segmentation, we achieve comparable results with UNETR which was optimized for segmentation. This reinforces our hypothesis that both tasks compliment and aid each other for a better performance. It also sheds light on how improving the segmentation task in turn hones the prognosis results and helps the model learn better representations of both images and EHR data.

\section{Conclusion and Future Work}
We propose an end-to-end multimodal framework for head and neck tumor diagnosis and prognosis in this work. This model takes advantage of the strengths of transformers in dealing with multimodal data and its ability to find long relations within and across modalities for better model performance. We train and validate our model on head and neck CT/PET images with patient EHR and compare our results with the current state-of-the-art methods for prognosis and segmentation.
For future work, self-supervised learning and pretraining of the network can be explored. They have proven to help models learn better, especially when the data is limited, as in our case. Additionally, the current network could be applied on similar tasks with different datasets to test the model for generalizability. 

\bibliographystyle{splncs04}
\bibliography{paper2727}

\begin{thebibliography}{10}
\providecommand{\url}[1]{\texttt{#1}}
\providecommand{\urlprefix}{URL }
\providecommand{\doi}[1]{https://doi.org/#1}

\bibitem{optuna}
Akiba, T., Sano, S., Yanase, T., Ohta, T., Koyama, M.: Optuna: {A}
  next-generation hyperparameter optimization framework. CoRR
  \textbf{abs/1907.10902} (2019), \url{http://arxiv.org/abs/1907.10902}

\bibitem{liver2}
Chen, J., Cheung, H., Milot, L., Martel, A.L.: Aminn: Autoencoder-based
  multiple instance neural network improves outcome prediction in multifocal
  liver metastases. In: International Conference on Medical Image Computing and
  Computer-Assisted Intervention. pp. 752--761. Springer (2021)

\bibitem{cox}
Cox, D.R.: Regression models and life-tables. Journal of the Royal Statistical
  Society: Series B (Methodological)  \textbf{34}(2),  187--202 (1972)

\bibitem{imagenet}
Deng, J., Dong, W., Socher, R., Li, L.J., Li, K., Fei-Fei, L.: Imagenet: A
  large-scale hierarchical image database. In: 2009 IEEE conference on computer
  vision and pattern recognition. pp. 248--255. Ieee (2009)

\bibitem{diamant}
Diamant, A., Chatterjee, A., Valli{\`e}res, M., Shenouda, G., Seuntjens, J.:
  Deep learning in head \& neck cancer outcome prediction. Scientific reports
  \textbf{9}(1),  1--10 (2019)

\bibitem{lung}
Doppalapudi, S., Qiu, R.G., Badr, Y.: Lung cancer survival period prediction
  and understanding: Deep learning approaches. International Journal of Medical
  Informatics  \textbf{148},  104371 (2021)

\bibitem{vit}
Dosovitskiy, A., Beyer, L., Kolesnikov, A., Weissenborn, D., Zhai, X.,
  Unterthiner, T., Dehghani, M., Minderer, M., Heigold, G., Gelly, S., et~al.:
  An image is worth 16x16 words: Transformers for image recognition at scale.
  arXiv preprint arXiv:2010.11929  (2020)

\bibitem{fh}
FH, T., CYW, C., EYW, C.: Radiomics ai prediction for head and neck squamous
  cell carcinoma (hnscc) prognosis and recurrence with target volume approach.
  BJR| Open  \textbf{3},  20200073 (2021)

\bibitem{fujima}
Fujima, N., Andreu-Arasa, V.C., Meibom, S.K., Mercier, G.A., Truong, M.T.,
  Hirata, K., Yasuda, K., Kano, S., Homma, A., Kudo, K., et~al.: Prediction of
  the local treatment outcome in patients with oropharyngeal squamous cell
  carcinoma using deep learning analysis of pretreatment fdg-pet images. BMC
  cancer  \textbf{21}(1),  1--13 (2021)

\bibitem{breast2}
Gupta, N., Kaushik, B.N.: Prognosis and prediction of breast cancer using
  machine learning and ensemble-based training model. The Computer Journal
  (2021)

\bibitem{UNETR}
Hatamizadeh, A., Tang, Y., Nath, V., Yang, D., Myronenko, A., Landman, B.,
  Roth, H.R., Xu, D.: Unetr: Transformers for 3d medical image segmentation.
  In: Proceedings of the IEEE/CVF Winter Conference on Applications of Computer
  Vision. pp. 574--584 (2022)

\bibitem{resnet}
He, K., Zhang, X., Ren, S., Sun, J.: Deep residual learning for image
  recognition. In: Proceedings of the IEEE conference on computer vision and
  pattern recognition. pp. 770--778 (2016)

\bibitem{lung2}
Hosny, A., Parmar, C., Coroller, T.P., Grossmann, P., Zeleznik, R., Kumar, A.,
  Bussink, J., Gillies, R.J., Mak, R.H., Aerts, H.J.: Deep learning for lung
  cancer prognostication: a retrospective multi-cohort radiomics study. PLoS
  medicine  \textbf{15}(11),  e1002711 (2018)

\bibitem{johnson}
Johnson, D.E., Burtness, B., Leemans, C.R., Lui, V.W.Y., Bauman, J.E., Grandis,
  J.R.: Head and neck squamous cell carcinoma. Nature reviews Disease primers
  \textbf{6}(1),  1--22 (2020)

\bibitem{kazmierskithesis}
Kazmierski, M.: Machine Learning for Prognostic Modeling in Head and Neck
  Cancer Using Multimodal Data. Ph.D. thesis, University of Toronto (Canada)
  (2021)

\bibitem{alexnet}
Krizhevsky, A., Sutskever, I., Hinton, G.E.: Imagenet classification with deep
  convolutional neural networks. Advances in neural information processing
  systems  \textbf{25},  1097--1105 (2012)

\bibitem{liver}
Lee, H., Hong, H., Seong, J., Kim, J.S., Kim, J.: Survival prediction of liver
  cancer patients from ct images using deep learning and radiomic feature-based
  regression. In: Medical Imaging 2020: Computer-Aided Diagnosis. vol. 11314,
  p. 113143L. International Society for Optics and Photonics (2020)

\bibitem{rectal}
Li, H., Boimel, P., Janopaul-Naylor, J., Zhong, H., Xiao, Y., Ben-Josef, E.,
  Fan, Y.: Deep convolutional neural networks for imaging data based survival
  analysis of rectal cancer. In: 2019 IEEE 16th International Symposium on
  Biomedical Imaging (ISBI 2019). pp. 846--849. IEEE (2019)

\bibitem{impotanceProg}
Mackillop, W.J.: The importance of prognosis in cancer medicine. TNM Online
  (2003)

\bibitem{hecktor}
Oreiller, V., Andrearczyk, V., Jreige, M., Boughdad, S., Elhalawani, H.,
  Castelli, J., Valli{\`e}res, M., Zhu, S., Xie, J., Peng, Y., et~al.: Head and
  neck tumor segmentation in pet/ct: The hecktor challenge. Medical image
  analysis p. 102336 (2021)

\bibitem{saeed2022ensemble}
Saeed, N., Majzoub, R.A., Sobirov, I., Yaqub, M.: An ensemble approach for
  patient prognosis of head and neck tumor using multimodal data (2022)

\bibitem{sobirov2022automatic}
Sobirov, I., Nazarov, O., Alasmawi, H., Yaqub, M.: Automatic segmentation of
  head and neck tumor: How powerful transformers are? arXiv preprint
  arXiv:2201.06251  (2022)

\bibitem{breast}
Sun, D., Wang, M., Li, A.: A multimodal deep neural network for human breast
  cancer prognosis prediction by integrating multi-dimensional data. IEEE/ACM
  transactions on computational biology and bioinformatics  \textbf{16}(3),
  841--850 (2018)

\bibitem{brain}
Sun, L., Zhang, S., Chen, H., Luo, L.: Brain tumor segmentation and survival
  prediction using multimodal mri scans with deep learning. Frontiers in
  neuroscience  \textbf{13}, ~810 (2019)

\bibitem{goglenet}
Szegedy, C., Liu, W., Jia, Y., Sermanet, P., Reed, S., Anguelov, D., Erhan, D.,
  Vanhoucke, V., Rabinovich, A.: Going deeper with convolutions. In:
  Proceedings of the IEEE conference on computer vision and pattern
  recognition. pp.~1--9 (2015)

\bibitem{vale}
Vale-Silva, L.A., Rohr, K.: Long-term cancer survival prediction using
  multimodal deep learning. Scientific Reports  \textbf{11}(1),  1--12 (2021)

\bibitem{wang}
Wang, X., Li, B.b.: Deep learning in head and neck tumor multiomics diagnosis
  and analysis: Review of the literature. Frontiers in Genetics  \textbf{12},
  ~42 (2021). \doi{10.3389/fgene.2021.624820},
  \url{https://www.frontiersin.org/article/10.3389/fgene.2021.624820}

\bibitem{WHOCancer}
WHO: {Cancer, \url{https://www.who.int/news-room/fact-sheets/detail/cancer}.
  Last accessed 30 Jan 2022}

\bibitem{mtlr}
Yu, C.N., Greiner, R., Lin, H.C., Baracos, V.: Learning patient-specific cancer
  survival distributions as a sequence of dependent regressors. Advances in
  neural information processing systems  \textbf{24},  1845--1853 (2011)

\bibitem{liver3}
Zhen, S.h., Cheng, M., Tao, Y.b., Wang, Y.f., Juengpanich, S., Jiang, Z.y.,
  Jiang, Y.k., Yan, Y.y., Lu, W., Lue, J.m., et~al.: Deep learning for accurate
  diagnosis of liver tumor based on magnetic resonance imaging and clinical
  data. Frontiers in oncology  \textbf{10}, ~680 (2020)

\bibitem{zhou}
Zhou, T., Fu, H., Zhang, Y., Zhang, C., Lu, X., Shen, J., Shao, L.: M2net:
  Multi-modal multi-channel network for overall survival time prediction of
  brain tumor patients. In: International Conference on Medical Image Computing
  and Computer-Assisted Intervention. pp. 221--231. Springer (2020)

\end{thebibliography}

\end{document}